# 'Black' TiO$_2$ nanotubes formed by high energy proton implantation show noble-metal-co-catalyst free photocatalytic H$_2$-evolution

*Ning Liu[1], Volker Häublein[2], Xuemei Zhou[1], Umamaheswari Venkatesan[3], Martin Hartmann[3], Mirza Mačković[4], Tomohiko Nakajima[5], Erdmann Spiecker[4], Andres Osvet[6], Lothar Frey[2], Patrik Schmuki[1]\**

[1]Department of Materials Science WW-4, LKO, University of Erlangen-Nuremberg, Martensstrasse 7, 91058 Erlangen, Germany;

[2]Fraunhofer Institute for Integrated Systems and Device Technology IISB, Schttkystrasse 10, 91058 Erlangen, Germany;

[3]ECRC - Erlangen Catalysis Resource Center, University of Erlangen-Nuremberg, Egerlandstrasse 3, 91058 Erlangen, Germany;

[4] Institute of Micro- and Nanostructure Research (WW9) & Center for Nanoanalysis and Electron Microscopy (CENEM), University of Erlangen-Nuremberg, Cauerstrasse 6, 91058 Erlangen, Germany;

[5]National Institute of Advanced Industrial Science and Technology, Tsukuba Central 5, 1-1-1 Higashi, Tsukuba, Ibaraki 305-8565, Japan;

[6]Department of Materials Sciences 6, iMEET, University of Erlangen-Nuremberg, Martensstrasse 7, 91058 Erlangen, Germany;

*Corresponding author. Tel.: +49 91318517575, fax: +49 9131 852 7582

Email: schmuki@ww.uni-erlangen.de

**Abstract:**

We apply high-energy proton ion-implantation to modify $TiO_2$ nanotubes selectively at their tops. In the proton-implanted region we observe the creation of intrinsic co-catalytic centers for photocatalytic $H_2$-evolution. We find proton implantation to induce specific defects and a characteristic modification of the electronic properties not only in nanotubes but also on anatase single crystal (001) surfaces. Nevertheless, for $TiO_2$ nanotubes a strong synergetic effect between implanted region (catalyst) and implant-free tube segment (absorber) can be obtained.

Ever since 1972, when Honda and Fujishima introduced photolysis of water using a single crystal of $TiO_2$, photocatalytic water splitting has become one of the most investigated scientific topics of our century [1]. The concept is strikingly simple: light (preferably sunlight) is absorbed in a suitable semiconductor and thereby generates electron-hole pairs. These charge carriers migrate in valence and conduction bands to the semiconductor surface where they react with water to form $O_2$ and $H_2$, respectively. Thus hydrogen, the energy carrier of the future, could be produced using just water and sunlight.

Key factors for an optimized conversion of water to $H_2$ are i) as complete as possible absorption of solar light (small band gap) while ii) still maintaining the thermodynamic driving force for water splitting (sufficiently large band-gap), including suitable band-edge positions relative to the water red-ox potentials, and iii) possibly most challenging – a sufficiently fast carrier transfer from semiconductor to water to obtain a reasonable reaction kinetics as opposed to carrier recombination or photo-corrosion [2-7].

In spite of virtually countless investigations on a wide range of semiconductor materials that in many respects are superior to titania (mostly in view of solar light absorption and carrier transport), $TiO_2$ still remains one of the most investigated photocatalysts. This is only partially due to suitable energetics but more so because of its outstanding (photo-corrosion) stability [2-7].

In general, the main drawbacks of $TiO_2$ are on the one hand its too large band-gap of 3-3.2 eV that allow only for about 7% of solar light absorption, and on the other hand that although a charge transfer to aqueous electrolytes is thermodynamically possible, the kinetics of these processes at the $TiO_2$/water interface are extremely slow if no suitable co-catalysts such as Pt, Au, Pd or similar are used [8-10].

However, in view of the first challenge (the 'too' large optical band-gap for efficient sunlight absorption), the recent finding of 'black' TiO$_2$ by Chen and Mao [11] seems to partially overcome this issue and has thus attracted accordingly wide scientific interest. The authors produced this modified form of TiO$_2$ (that showed strong visible absorption) by exposing anatase TiO$_2$ nanoparticles to a high pressure/high temperature treatment in H$_2$. This material was found to show a range of outstanding functional features. In the original work, Chen and Mao demonstrated a significantly increased photocatalytic activity for water splitting when black TiO$_2$ was loaded with a Pt co-catalyst and used under bias-free conditions (i.e. used directly as a nanoparticle suspension in an aqueous/methanol solution under sunlight (AM 1.5) conditions). The high catalyst activity was attributed to a thin amorphous TiO$_2$ hydrogenated layer that was formed under high pressure treatment and that encapsulated the anatase core of the nanoparticles, leading to a considerable narrowing of the optical absorption band-gap of the treated material (turning its appearance to black).

Follow-up work mainly replaced the high pressure treatment by various other reductive treatments (high temperature Ar/H$_2$, Ar, vacuum exposure, or electrochemical reduction, etc. [12-16]) – such material was reported to cause a similar effect on the Pt-co-catalyzed photocatalytic H$_2$-production or when used in a range of other electrochemical applications [17-19].

More recently, we reported that high pressure/high temperature/hydrogen-treated TiO$_2$ (HPT-TiO$_2$) in form of nanotubes [20] and powder [21] shows another key feature, that is, a strongly enhanced photocatalytic activity for hydrogen production in absence of any noble metal co-catalyst [22]. The work thus showed that a HPT-treatment not only leads to the effects observed by Chen and Mao, but additionally suggested that a co-catalytic center in TiO$_2$ is formed – similar in its effect to noble metal decoration. This catalytic center was proposed to involve an unusually stable Ti$^{3+}$ species with characteristic EPR and PL features [20, 21].

Importantly, this noble-metal-free co-catalysis effect was found to be specific for HPT hydrogenation. Other treatments to form 'black titania', namely the above mentioned high temperature treatments of $TiO_2$ in other reductive environments, did not lead to any comparable co-catalytic center, neither for nanotubes [20] nor for nanoparticles [21].

In the present work, we show, however, that a so far unexplored technique – that is high energy proton implantation – can similarly activate this co-catalytic effect, not only in $TiO_2$ nanotubes but as well in anatase single crystals.

For our experiments, we used $TiO_2$ nanotube arrays of various length (1-12 μm) grown by self-organizing anodization and annealed to anatase, as well as single crystal anatase wafers with an epi-polished [001] surface (see SI for details). These substrates then were ion implanted with protons ($p^+$) at an energy of 30 keV with a dose of $10^{16}$ ions/cm$^2$ using a Varian 350 D ion implanter (more details are given in the SI). The implanted (and corresponding non-implanted reference samples) were then tested for their $H_2$ evolution activity under AM1.5, 100 mW/cm$^2$ light (details are given in the experimental section). The results in Fig. 1a show that in every case - that is for all nanotube lengths as well as on the single crystal - $p^+$-implantation leads to a modification that becomes active for photocatalytic $H_2$ evolution. While the amount of hydrogen produced on the implanted [001] crystal is relatively low, for the tubes a clear increase with the tube length can be observed. Fig. 1b shows a calculated ion and damage depth distribution for $TiO_2$ nanotubes according to Monte-Carlo simulation using TRIM 2008 [23]. Fig. S1 SI gives the corresponding depth distributions for compact anatase. For the single crystal (Fig. S1 SI), implantation leads to an implant/damage zone reaching to approx. 350 nm below the surface with a maximum of implanted H-ions at approx. 0.25 μm below the surface and maximum lattice damage zone (vacancies/interstitials) peaking at approx. 0.2 μm.

For tubes (Fig. 1b), due to their open volume (simulated as porosity) the implant zone extends to approx. 1.2 μm tube length with a maximum of H implanted in a depth of ~900 nm and a maximum in lattice defects at a depth of ~800 nm. In every case, for single crystal and nanotubes, ion implantation did not cause any change in morphology that could be detected by SEM – an example of the tube morphology in SEM before and after implantation is shown in Fig. 1c.

If the implant/damage distributions in Fig. 1b are compared with the photoactivity of the different tube lengths, it becomes clear that the resulting hydrogen evolution efficiency is not directly correlated to the implant amount or relative distribution (for all tubes only the top 1 μm is affected with the same dose, while the rest of the tube remains relatively defect free). One can thus ascribe the length effect to deeper light penetration into the tubes and carrier generation in the underlying defect free zone. For light of an energy around the band-gap of $TiO_2$ (~3 eV), light penetrates several micrometers into $TiO_2$ nanotubes [24]. That is, for longer tubes the length for absorption increases and excited electrons generated in this zone can reach the activated tube tops by diffusion. (The electron diffusion length in $TiO_2$ nanotubes has been reported to be several 10 micrometers [25], and thus seems not to represent a limiting factor in the present case.)

To characterize the structural and morphological changes induced by H-implantation we used XRD (Fig. 2a and 2b) and transmission electron microscopy (TEM) techniques (Fig. 2c-h). XRD and selected area electron diffraction (SAED) were measured for a 1 μm long tube layer (i.e. averaging the information over the entire tube length, using tubes fully affected by H-implantation). High-resolution TEM (HRTEM) (Fig. 2c and 2d) and bright-field (BF) TEM (Fig. 2g and 2h) imaging was carried out in the top part of implanted and reference tubes.

XRD and corresponding Rietveld refinement (Fig. 2a and 2b) show for the non-implanted reference sample the typical anatase pattern of conventionally annealed $TiO_2$ nanotubes [26]. After implantation, clearly a significant decrease of the anatase peak intensities and broadening of peaks can be observed which indicates reduction of the length of structural coherence (usually due to amorphization and/or discretization of crystallites). In the reference nanotubes, the diffraction pattern showed weak preferential (001)-orientation. The Lotgering factor $F$(001) was calculated to be 49.2%. However, it turned to be only 2.2% after $p^+$-implantation. Observed peak broadening and large reduction of orientation degree indicate that the H-implantation efficiently interacts with the $TiO_2$ lattice.

HRTEM images of reference and H-implanted nanotubes are shown in Figure 2c and 2d, respectively. In both HRTEM images lattice spacings of 0.35 nm are observed, which fit well to (101) lattice planes of anatase. Also some randomly oriented nanocrystallites with various shapes and sizes are visible in Figure 2d. HRTEM reveals also the presence of thin amorphous rims around the samples (as indicated with white arrows and circles in Fig. 2c and d, and SI Fig. S3, S4), i.e. this feature is present for non-implanted and implanted samples (and is often observed for HRTEM images of nanoparticle samples). Nevertheless, the amount of amorphous material is higher for the proton-implanted samples. This is not only in line with the XRD data, but also apparent from the SAED patterns (see Figure 2e and 2f). In both cases nanocrystalline anatase ($TiO_2$ with a tetragonal crystal structure is present, which is also consistent with ICSD 9852). However, the SAED pattern of nanotubes after $p^+$-implantation exhibits a blurred halo (background signal), in addition to the regular diffraction rings. This indicates a presence of a higher amount of amorphous species in the vicinity of the nanocrystallites. Furthermore, and in particular visible for the (101) reflections in the SAED patterns (compare Figure 2e and 2f), a broadening of the diffraction rings is observed, which further confirms a reduced average crystallite size in the nanotubes after $p^+$-implantation.

The changes induced by implantation are particularly apparent from BF TEM images taken under defocus (Fresnel contrast) conditions (Fig. 2g and 2h). Clearly H-implantation causes an increased amount of voids and thus a reduction of the crystallite size, which is in line with XRD and the broadening of the diffraction rings in the SAED pattern shown in Fig. 2f. Such voids, after proton implantation, are a common observation [27-31] and can be ascribed to point defects (vacancies or interstitials) diffusion and agglomeration. In our case, the voids in the implanted material discretize the wall on length scale of around 10-20 nm. TEM data before implantation under defocus conditions (Fig. 2g) show the presence of square facets. This faceting after implantation disappears (in Fig. 2h). This behavior is consistent with XRD that shows a clear reduction of the orientation degree.

Additionally, Raman spectra were acquired for the implanted/non-implanted single crystal samples and compared to the tube response (see SI Fig. S5 and S6). Notable is particularly in the single crystal after implantation a relative increase of the Eg bands at 144 and 636 cm$^{-1}$ relative to the $B_{1g}$ and the $A_{1g}$ peaks. This indicates a break of the symmetry of the (001) plane [32]. For the tubes, the main effect is a mild blue shift, evident e.g. at the main Eg peak, that is at 144 cm$^{-1}$ for the non-implanted tubes to 146 cm$^{-1}$ for the implanted material and a widening of the FWHM from 13 to 16 cm$^{-1}$. These observations are in line with models that indicate phonon confinement by a decrease of the effective particle size. In fact, if various models for phonon confinement [33-35] (see SI) are applied to the shifts in peak position and the changes in the FWHM at the Eg peak, the corresponding length scale results as ~10-20 nm − this well in line with direct observation of voids in the TEM walls with spacings at ≈ 20 nm (Fig. 2h).

In order to gain additional information on the electronic nature of the defect structure, we carried out photoluminescence (PL) and electron paramagnetic resonance (EPR) measurements. Fig. 3a shows that distinct changes are apparent in the PL response of the tubes due to p$^+$-implantation. While the non-implanted reference tube shows a PL peaking at 600-700 nm, which is typical for anatase TiO$_2$ and has been widely ascribed to the recombination of self-trapped excitons, the PL from the H-implanted nanotube sample shows a shift of the maximum and a strong tail towards the higher energies. I.e. this PL is likely originating from emissive recombination states close to the conduction band of anatase. For the single crystal (inset Fig. 3a), the main effect of the implantation is a strong enhancement in the PL intensity peaking at ~ 660 nm to ~ 570 nm, which again reflects a drastic increase in emissive recombination states closer to the conduction band induced by ion implantation and corresponding damage.

From EPR spectra taken for implanted and reference tubes at 90 K in the dark and under illumination, a different defect signature becomes apparent (Fig. 3b). The dark conditions indicate the presence of an additional paramagnetic defect in the implanted samples. This feature becomes even more visible under illumination. This Ti$^{3+}$ center can be described by orthorhombic g values [1.990 1.929 1.909], with a considerable distribution in the g strain contribution which is significantly different from the typically observed Ti$^{3+}$ centers found in the reduced titania [36, 37].

Overall the PL and EPR findings strongly support that a main electronic effect of p$^+$-implantation is the creation of paramagnetically active recombination states that are close to the band gap of the anatase. Such features were also observed for high pressure hydrogenated anatase, where they have been identified as a key indicator for establishing intrinsic H$_2$ evolution ability in TiO$_2$ [20, 21].

In conclusion, the present work shows that high-energy proton implantation in anatase is able to form an intrinsic co-catalytic center for photocatalytic H$_2$ evolution. While the ion implantation into a (001) surface plane of an anatase crystal leads to a comparably low H$_2$ production efficiency, implantation into TiO$_2$ nanotubes provides a remarkable length effect.

I.e. clearly a synergistic interaction between the implanted and the intact tube segments is observed. This may be attributed to a coupling between the intact lower tube part as light absorber and the catalytically active center in the upper tube part, where conduction band electrons are directed from the lower tube part to the active tube tips. This effect is suppressed in the single crystal, as in this case holes generated in the underlying intact part cannot be captured by the electrolyte (due to the short diffusion length of $h^+$ in $TiO_2$ [25]). The effects of the proton implantation as seen from TEM, XRD and Raman measurements are the creation of structural damage sites (vacancy/interstitial pairs), and a reduction of the length of structural coherence (amorphization, void formations, and release of texture). In PL and EPR, characteristic defect signatures (states close to the conduction band) are observed. These results resemble surface modifications observed for 'black titania' (tubes or powder) formed by high-pressured hydrogenation, and thus further show that suitable structural defect engineering can effectively activate anatase $TiO_2$ nanotubes or single crystals for photocatalytic noble metal free $H_2$ generation.

SUPPORTING INFORAMTION:

Experimental details on the fabrication.

Additional TRIM calculation depth distribution of proton implantation in $TiO_2$ single crystal (Fig. S1).

Additional TEM and corresponding electron diffraction patterns (Fig. S2-S4)

Raman spectra for (001) anatase single crystal and $TiO_2$ nanotubes before and after H-implantation (Fig. S5, S6).

Calculation models and reference papers for the Eg Raman line shift and FWHM as a function of $TiO_2$ feature size (Fig. S7, S8).

Calculation details for Rietveld analysis.

ACKNOWLEDGMENT

We would like to acknowledge the ERC, the DFG, and the Erlangen DFG cluster of excellence (EAM) for financial support and Yuyun Yang for Raman measurements.

**Figure captions:**

**Fig. 1** a) Photocatalytic $H_2$ production under open circuit conditions in methanol/water (50/50 vol%) of $TiO_2$ nanotube layers of different thickness illumination before and after H-implantation (measured under AM 1.5, 100 mW/cm$^2$) (inset: photocatalytic $H_2$ production of (001) single crystal anatase before and after H-implantation), (grey box represent no detected hydrogen evolution; red box represents detectable amount of hydrogen evolution); b) Calculation depth distribution of implanted ions (H ions) and crystal damage (Ti-, Orecoil-) for $TiO_2$ nanotubes; c) SEM images of $TiO_2$ nanotube layer before and after H-implantation.

**Fig. 2** X-ray diffraction spectra (XRD) and Rietveld refinement of $TiO_2$ nanotubes before (a) and after (b) implantation; HRTEM images (c, d) and inverted SAED patterns (e, f) of $TiO_2$ nanotubes before and after H-implantation, respectively (white circles indicate the amorphous regions). Insets in the SAED patterns are intensity profiles which are obtained by radial averaging the respective diffraction patterns; BF TEM images taken under focus (left), underfocus (center) and overfocus (right) conditions for $TiO_2$ nanotubes before (g) and after (h) implantation, showing characteristic Fresnel contrast indicating the presence of voids. The white and black arrows indicate exemplarily voids, which become visible in under- and overfocus.

**Fig. 3** a) Photoluminescence of $TiO_2$ nanotubes and (001) single crystal (inset) before and after implantation measured in air using 375 nm excitation; b) EPR spectra for $TiO_2$ nanotubes before and after H-implantation at 90 K in dark after 90 min illumination. The simulations are in green lines for HIM tubes: g values [1.990 1.929 1.909], g strain [0.020 0.075 0.100].

**Fig. 1a**

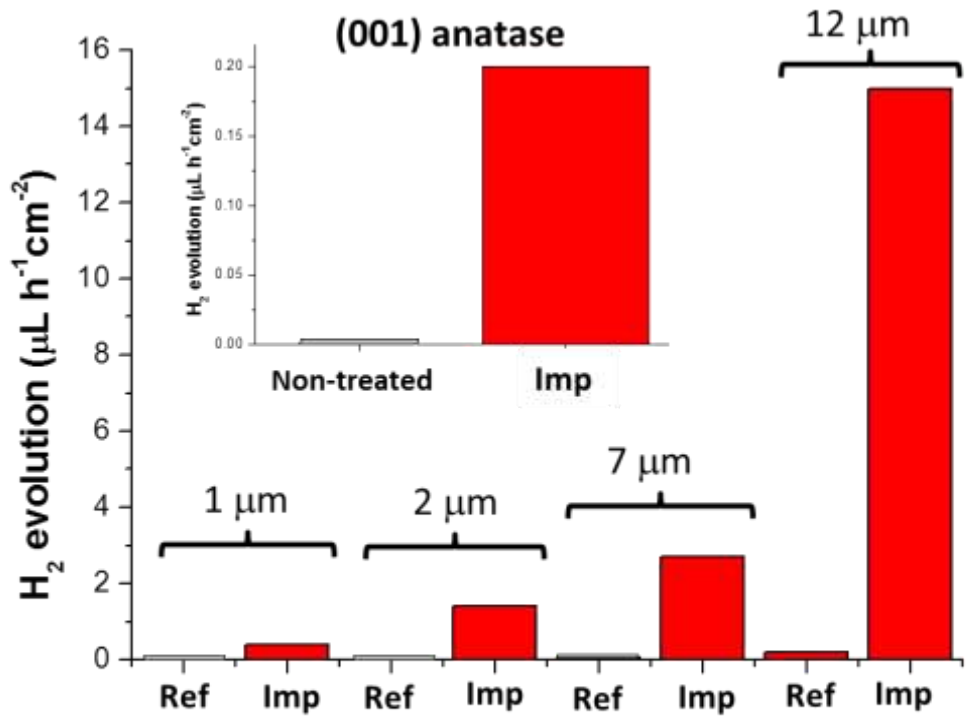

**Fig. 1b**

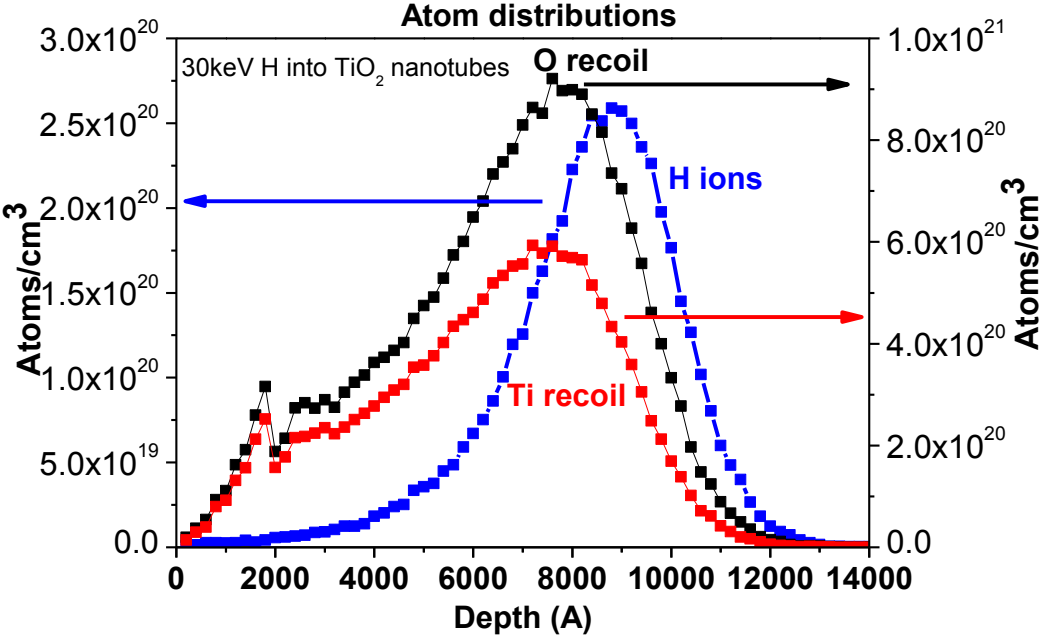

**Fig. 1c**

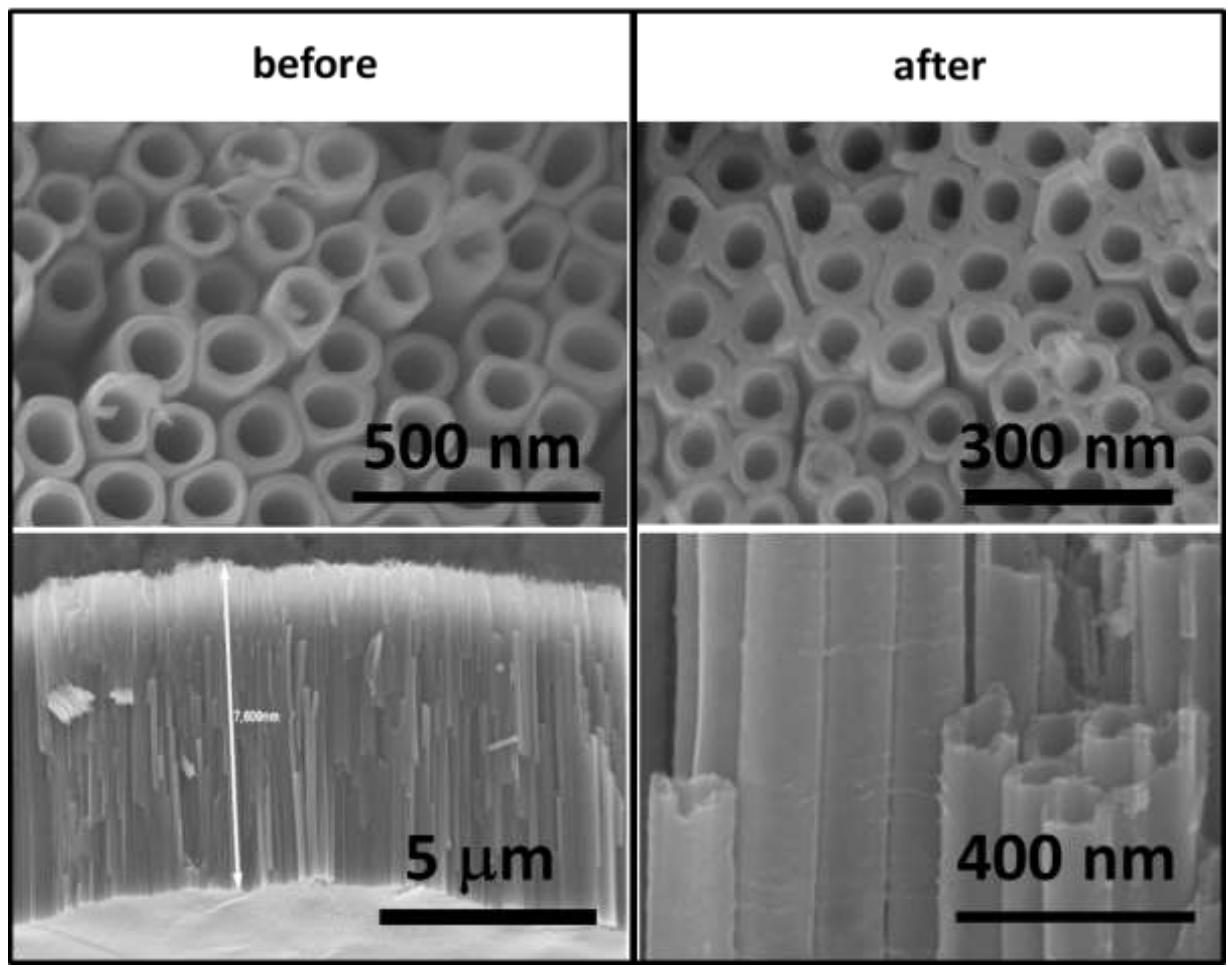

**Fig. 2**

(a)
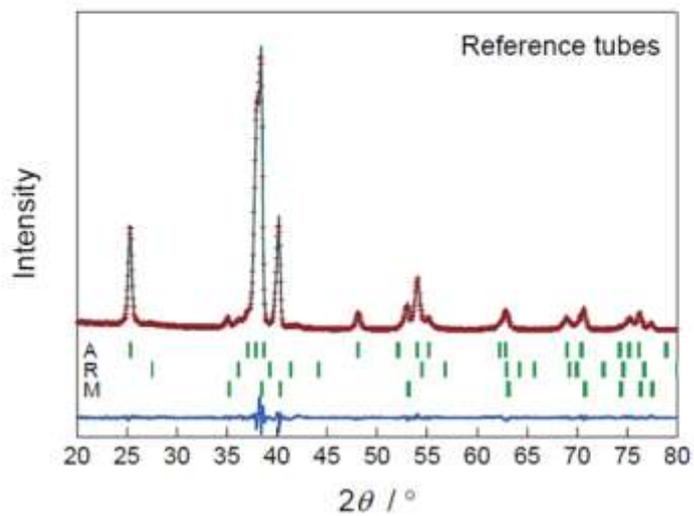

(b)
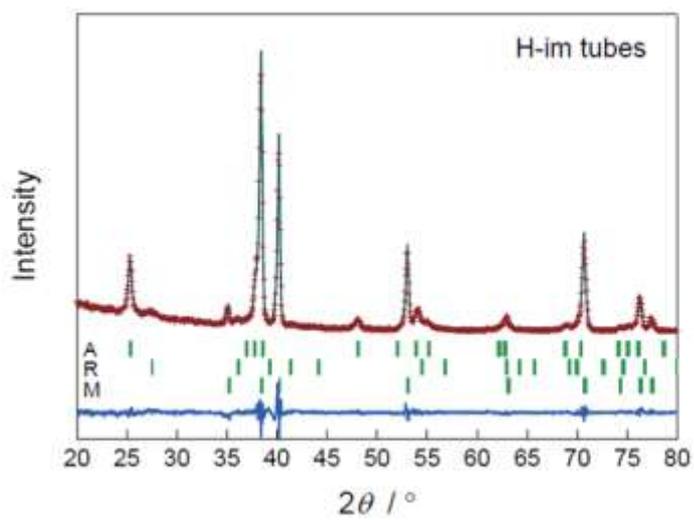

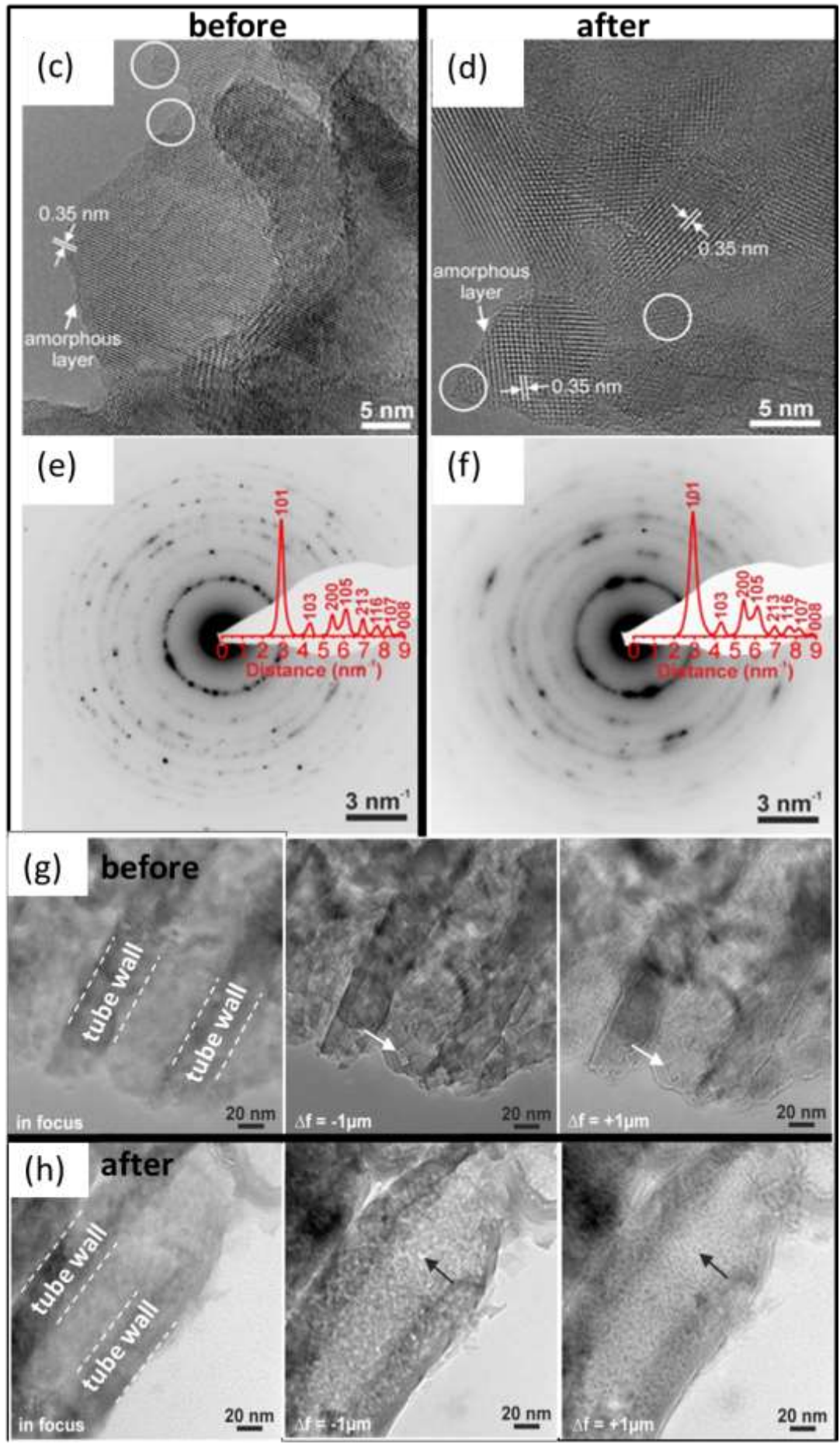

Fig. 2

Fig. 3a

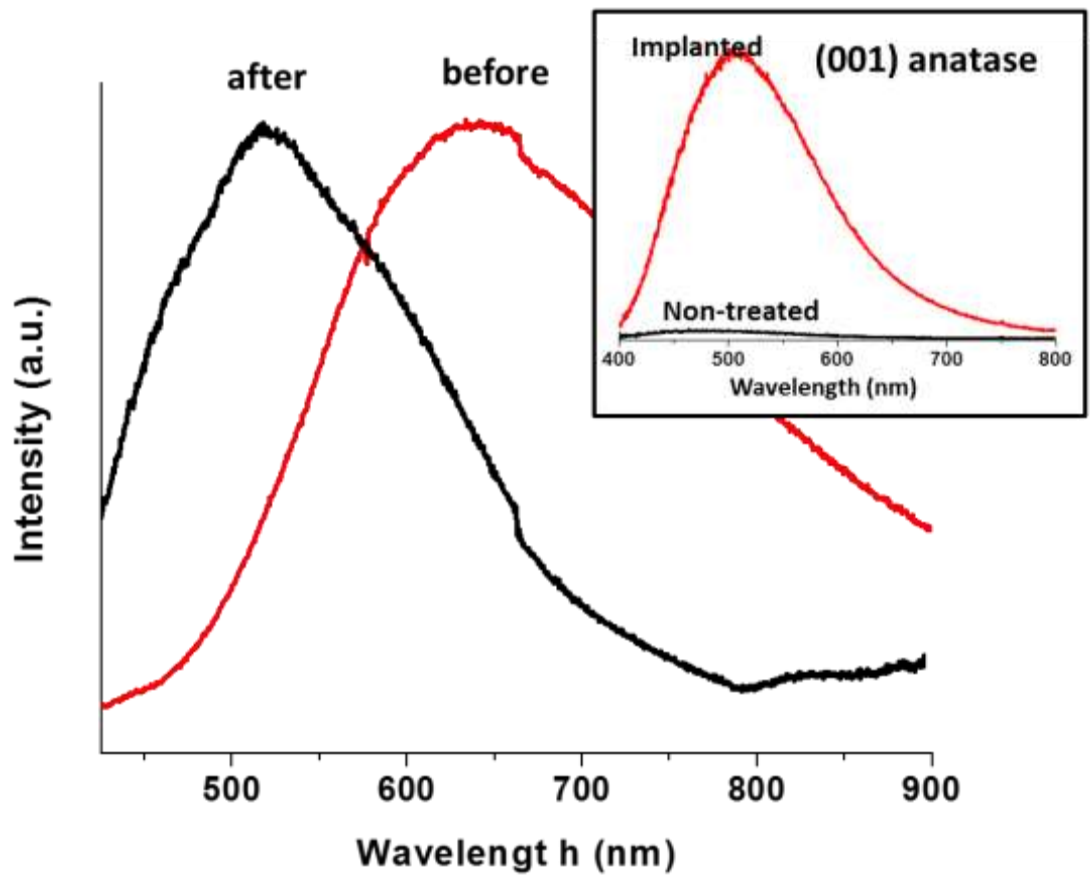

**Fig. 3b**

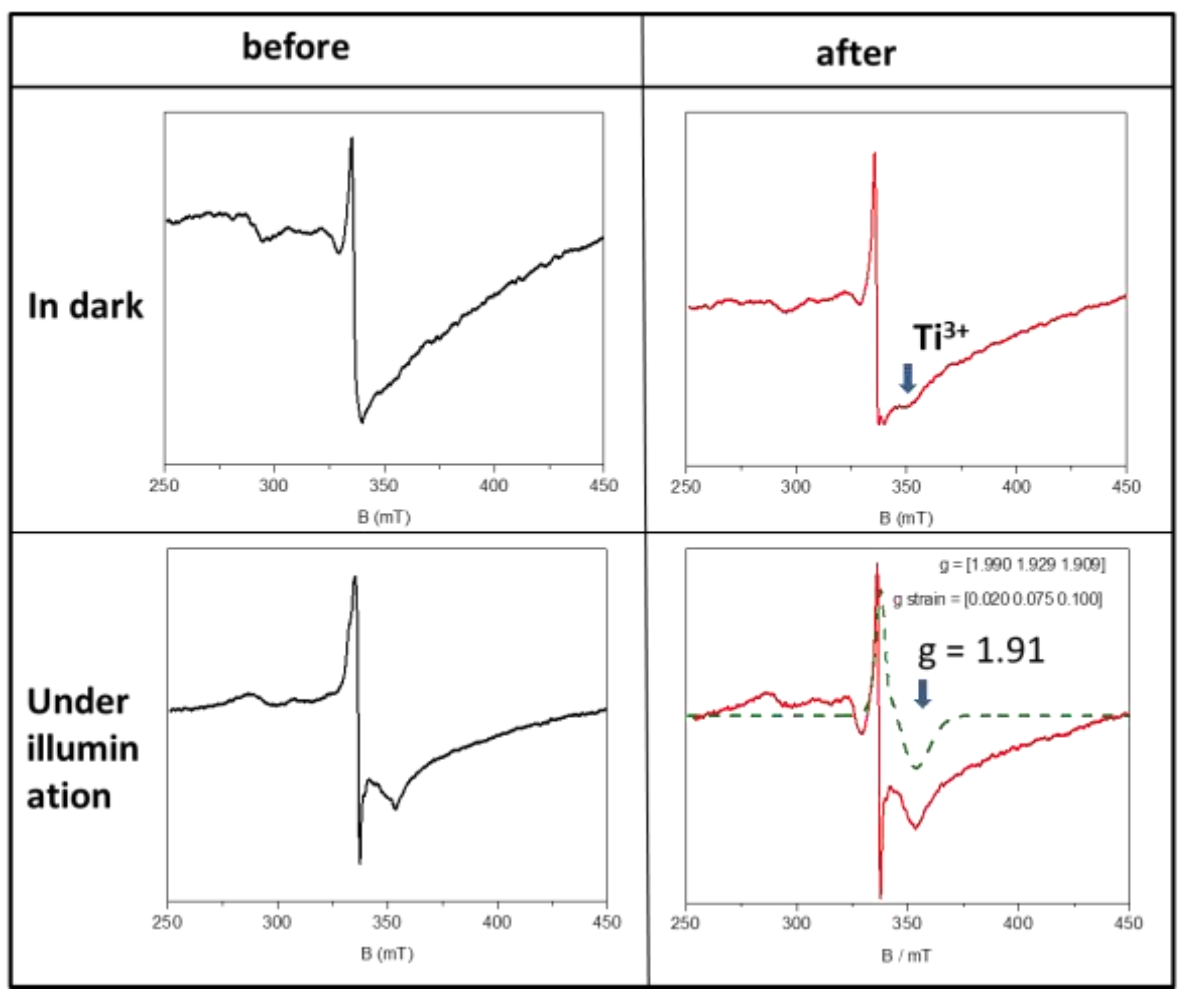

**Table of Contents**

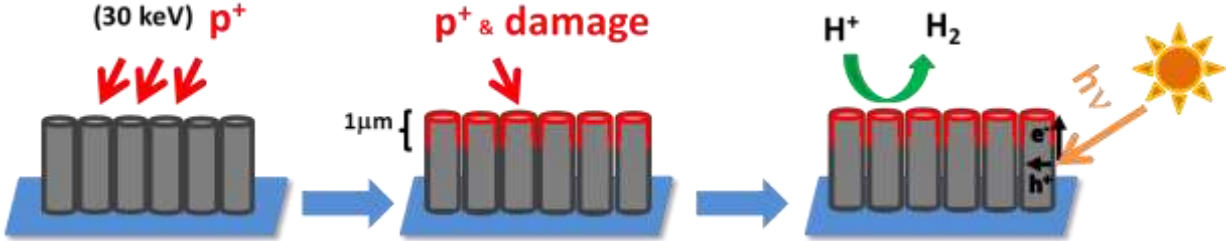

# Supporting information

# 'Black' TiO$_2$ nanotubes formed by high energy proton implantation show noble-metal-co-catalyst free photocatalytic H$_2$-evolution

*Ning Liu[1], Volker Häublein[2], Xuemei Zhou[1], Umamaheswari Venkatesan[3], Martin Hartmann[3], Mirza Mačković[4], Tomohiko Nakajima[5], Erdmann Spiecker[4], Andres Osvet[6], Lothar Frey[2], Patrik Schmuki[1]\**

[1]Department of Materials Science WW-4, LKO, University of Erlangen-Nuremberg, Martensstrasse 7, 91058 Erlangen, Germany;

[2]Fraunhofer Institute for Integrated Systems and Device Technology IISB, Schttkystrasse 10, 91058 Erlangen, Germany;

[3]ECRC - Erlangen Catalysis Resource Center, University of Erlangen-Nuremberg, Egerlandstrasse 3, 91058 Erlangen, Germany;

[4] Institute of Micro- and Nanostructure Research (WW9) & Center for Nanoanalysis and Electron Microscopy (CENEM), University of Erlangen-Nuremberg, Cauerstrasse 6, 91058 Erlangen, Germany;

[5]National Institute of Advanced Industrial Science and Technology, Tsukuba Central 5, 1-1-1 Higashi, Tsukuba, Ibaraki 305-8565, Japan;

[6]Department of Materials Sciences 6, iMEET, University of Erlangen-Nuremberg, , Martensstrasse 7, 91058 Erlangen, Germany;

*Corresponding author. Tel.: +49 91318517575, fax: +49 9131 852 7582

Email: schmuki@ww.uni-erlangen.de

# Experimental:

As substrates for $TiO_2$ nanotube growth we used titanium foils (99.6% purity, Goodfellow) with a thickness of 0.1 mm. Prior to tube formation the foils were cleaned by sonication in acetone and ethanol followed by rinsing with deionized (DI) water and drying in a nitrogen stream. To perform electrochemical $TiO_2$ nanotube formation, the foils were anodized using a power supply (Voltcraft VLP 2403 pro) in a two electrode configuration with a counter electrode made from platinum gauze. The typical electrolyte for $TiO_2$ nanotubes was prepared from ethylene glycol (EG, Sigma–Aldrich, containing less than 0.2 wt% $H_2O$), with addition of 1 M DI $H_2O$ and 0.1 M $NH_4F$ (Sigma–Aldrich, 98%). The anodization was carried out at 60V for 2, 5, 15 and 30 min, and $TiO_2$ nanotube layers of a thickness of about 1, 2, 7 and 12 μm were obtained.

Thermal treatments of the nanotube layers were carried out in air using a Rapid Thermal Annealer (Jipelec JetFirst 100) at 500 °C with heating/cooling rates of 1 °C/s. The samples were annealed at 450 °C for 1 h.

The single crystal anatase wafers were obtained from natural anatase to an epi-polished (001) surface (SurfaceNet GmbH, Germany).

Proton implantation was carried out at an energy of 30 keV and a nominal dose of $10^{16}$ ions/cm$^2$ using a Varian 350 D ion implanter.

A Hitachi FE-SEM S4800 was used for morphological characterization of the samples. The length of the nanotubes was directly obtained from SEM cross-sections. XRD patterns were collected using an X'pert Philips PMD diffractometer with a Panalytical X'celerator detector, using graphite-monochromatized CuKα radiation ($\lambda$ = 1.54056Å). The chemical composition of the layer was characterized with X-ray photoelectron spectroscopy (XPS, PHI 5600 XPS spectrometer, US).

Transmission electron microscopy (TEM) was performed with a Titan$^3$ Themis 300, a Phillips CM300 UltraTWIN and a Philips CM30 TWIN/STEM (FEI Company, Netherlands). The Titan$^3$ Themis 300 is equipped with a high-brightness field-emission gun (X-FEG), a monochromator system (energy resolution 0.2 eV), two $C_s$-correctors (probe and image side) from CEOS (Corrected Electron Optical Systems GmbH), a Super-X detector (for energy dispersive X-ray spectroscopy), a Gatan Imaging Filter, a high-angle annular dark-field (HAADF) detector and a 4k CMOS camera. This microscope was operated at 200 kV acceleration voltage. The Philips CM300 UltraTWIN and the CM30 TWIN/STEM microscopes are equipped with $LaB_6$ filaments, 2k and 1k charged coupled device cameras from TVIPS (Germany), respectively, and were operated at 300 kV acceleration voltage. For TEM analysis the $TiO_2$ nanotubes are prepared on commonly used copper TEM grids coated with a holey carbon film. During TEM analysis no noticeable electron-beam-induced damage was observed. The free available software ImageJ (version 1.48r) and the commercially available software

DigitalMicrograph[TM] were used for image analysis. The evaluation of the electron diffraction patterns was performed by using the software JEMS[1] (version version 3.7624U2012) and the inorganic crystal structure database (ICSD).

The room temperature CW EPR spectra were recorded on an X-band ($\nu_{mw}$ = 9.84 GHz) EMXmicro BRUKER spectrometer and at 70 K using an Oxford flow cryostat with liquid nitrogen flow. The $B_0$ modulation amplitude used was 0.4 mT, and the modulation frequency was adjusted to $\nu_{mod}$ = 100 kHz. The microwave power used was low enough to prevent the saturation of the spin systems.

The photoluminescence (PL) of the powder samples was excited with a 375 nm diode laser and the spectra were recorded at room temperature with an iHR320 monochromator and Synergy Si CCD camera (both Horiba Jobin-Yvon). The spectra are corrected for the spectral sensitivity of the setup, determined with the help of a calibrated halogen lamp.

Measurements of Raman spectra were performed on a Spex 1403 Raman Spectrometer. A line (632 nm) of a HeNe laser was taken as the excitation source.

Photocatalytic hydrogen generation was measured under open circuit conditions from an aqueous methanol solution (50 vol%) under AM 1.5 (100 mW/cm$^2$) solar simulator illumination. The amount of $H_2$ produced was measured using a Varian gas chromatograph with a TCD detector. For rate determination, data were taken approximately every 24 h during solar simulator irradiation. To prepare suspensions for $H_2$ measurements, 2 mg $TiO_2$ powders were dispersed in 10 mL of DI water/methanol (50/50 v%) with ultrasonication for 30 min. During illumination, the suspensions were continuously stirred.

Monte-Carlo simulations of the implant and damage depth-distributions were carried out using TRIM 2008 and 2013 [2]. (We consider the small peaks appearing in the profiles at energies ~ 180 nm (tube) and 50 nm (single crystal) as artifacts of the TRIM code).

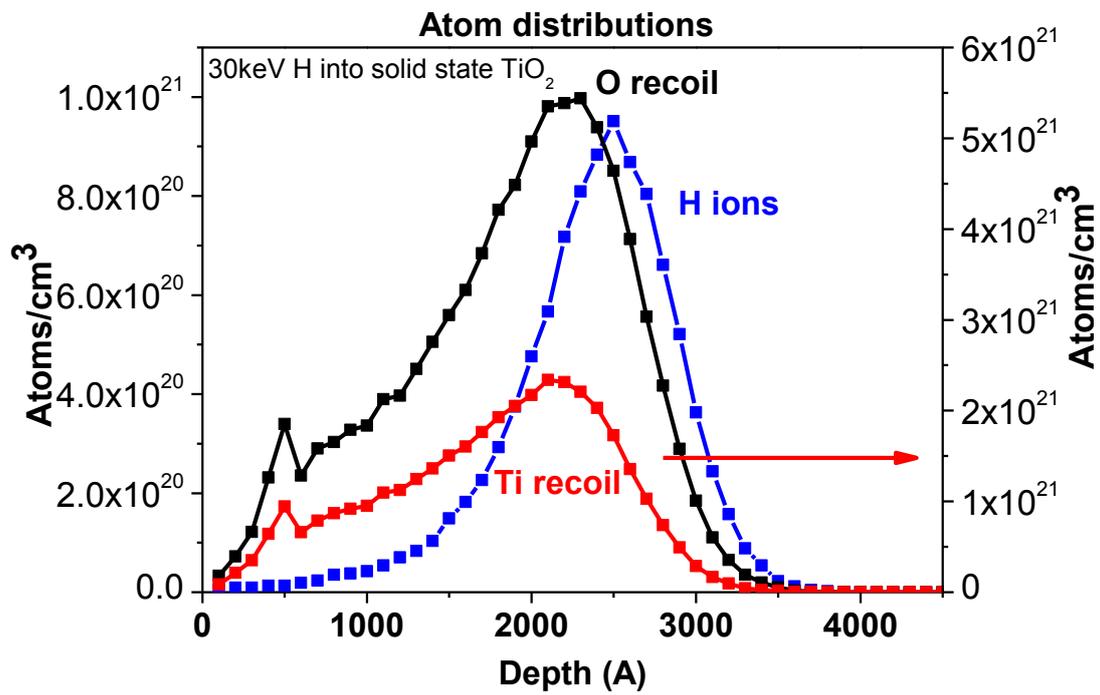

**Fig. S1** Calculation depth distribution of implanted protons (H ions) and crystal damage (titanium and oxygen recoil) in pure $TiO_2$ anatase substrate.

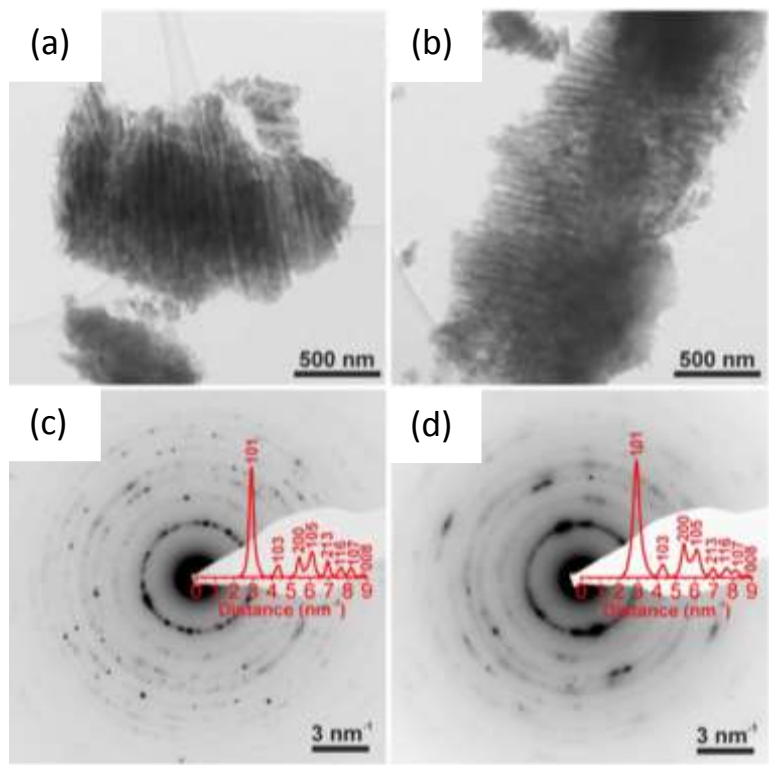

**Fig. S2** Representative bright-field (BF) TEM images in Figure 1a) and b) show bundles of $TiO_2$ nanotubes after conventional annealing in air (to anatase) and after H implantation, respectively and corresponding electron diffraction patterns (see Fig. 2c and d).

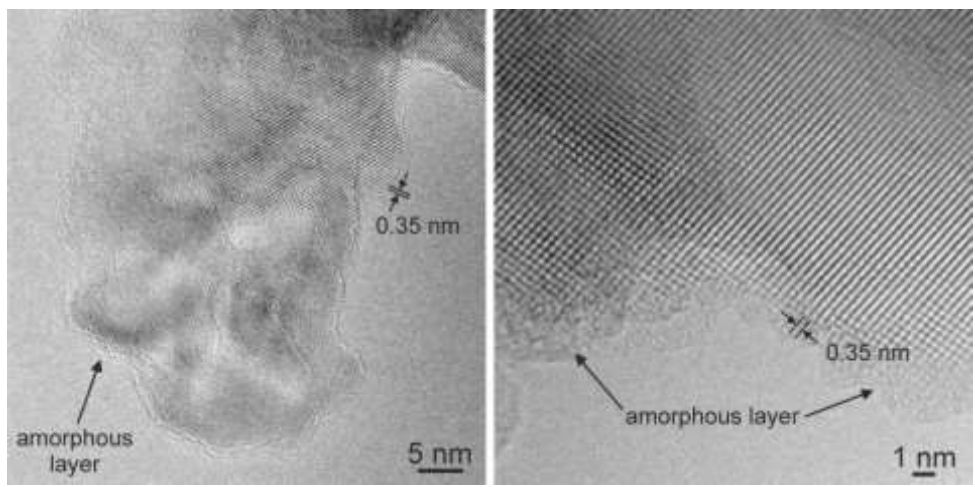

**Figure S3.** HRTEM images of reference TiO$_2$ nanotubes.

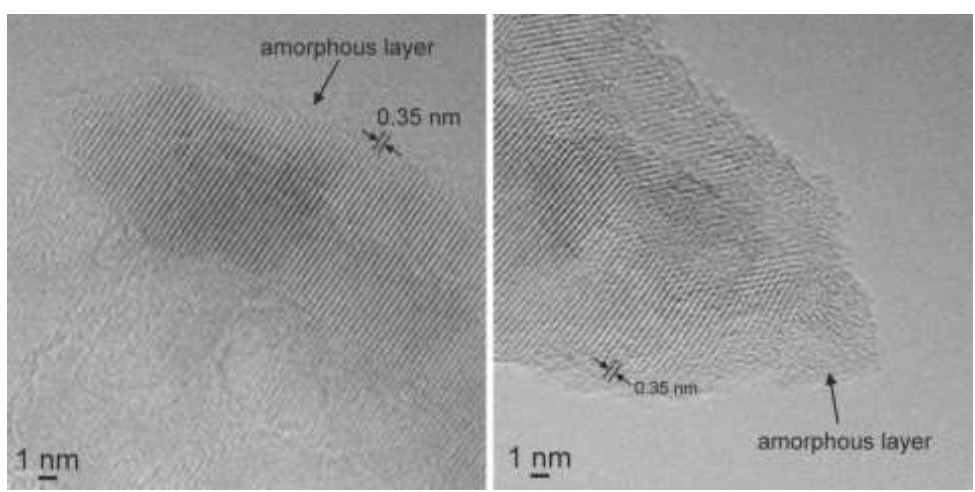

**Figure S4.** HRTEM images of TiO$_2$ nanotubes after H-implantation.

**Fig. S3 and Fig. S4** Additional HRTEM images of samples are provided, which support statements given in the manuscript. A careful analyses under HRTEM conditions also revealed that electron beam irradiation does not introduce effects, such as re-crystallization, amorphization and/or pore formation in the nanotubes. These images indicate that amorphous layers are present in as annealed and implanted tubes.

**Raman Spectra:**

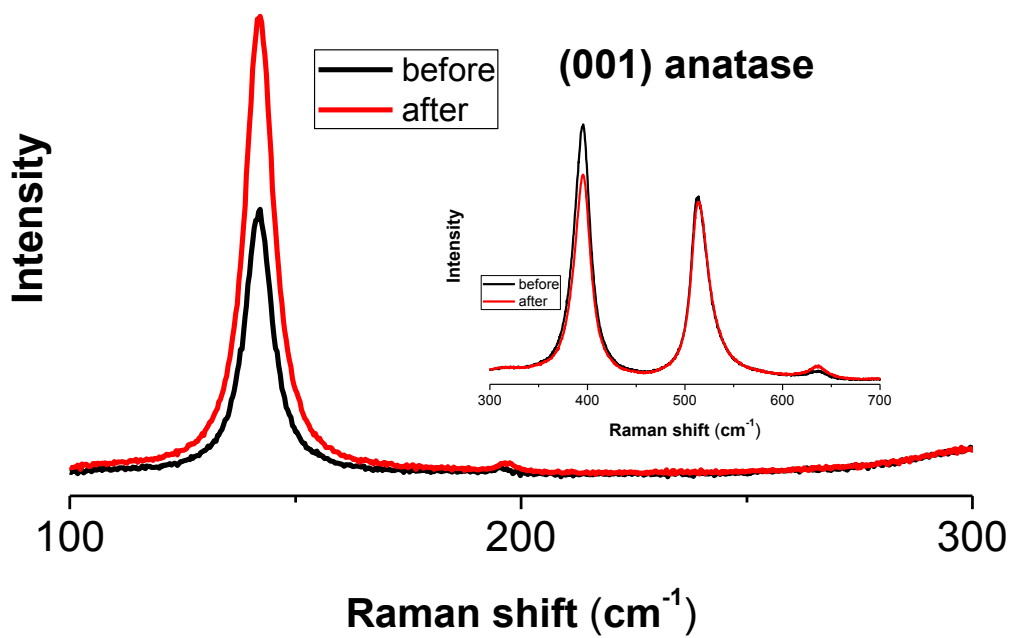

**Fig. S5** Raman spectra of (001) anatase single crystal before and after H-implantation.

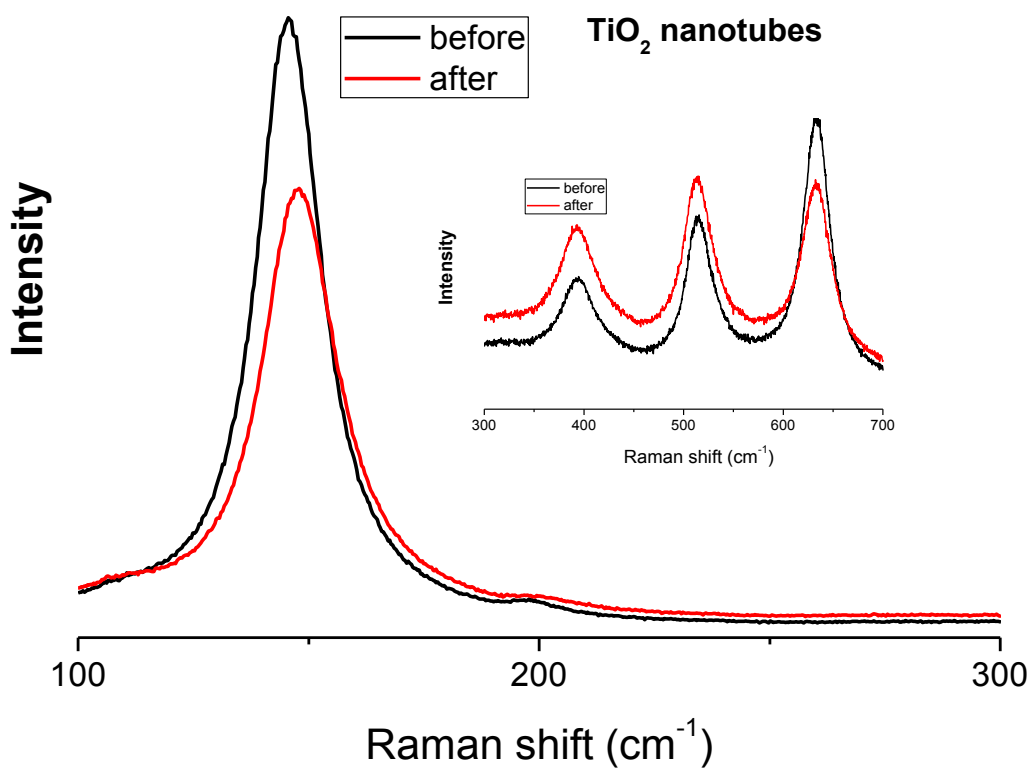

**Fig. S6** Raman spectra for $TiO_2$ nanotubes before and after H-implantation.

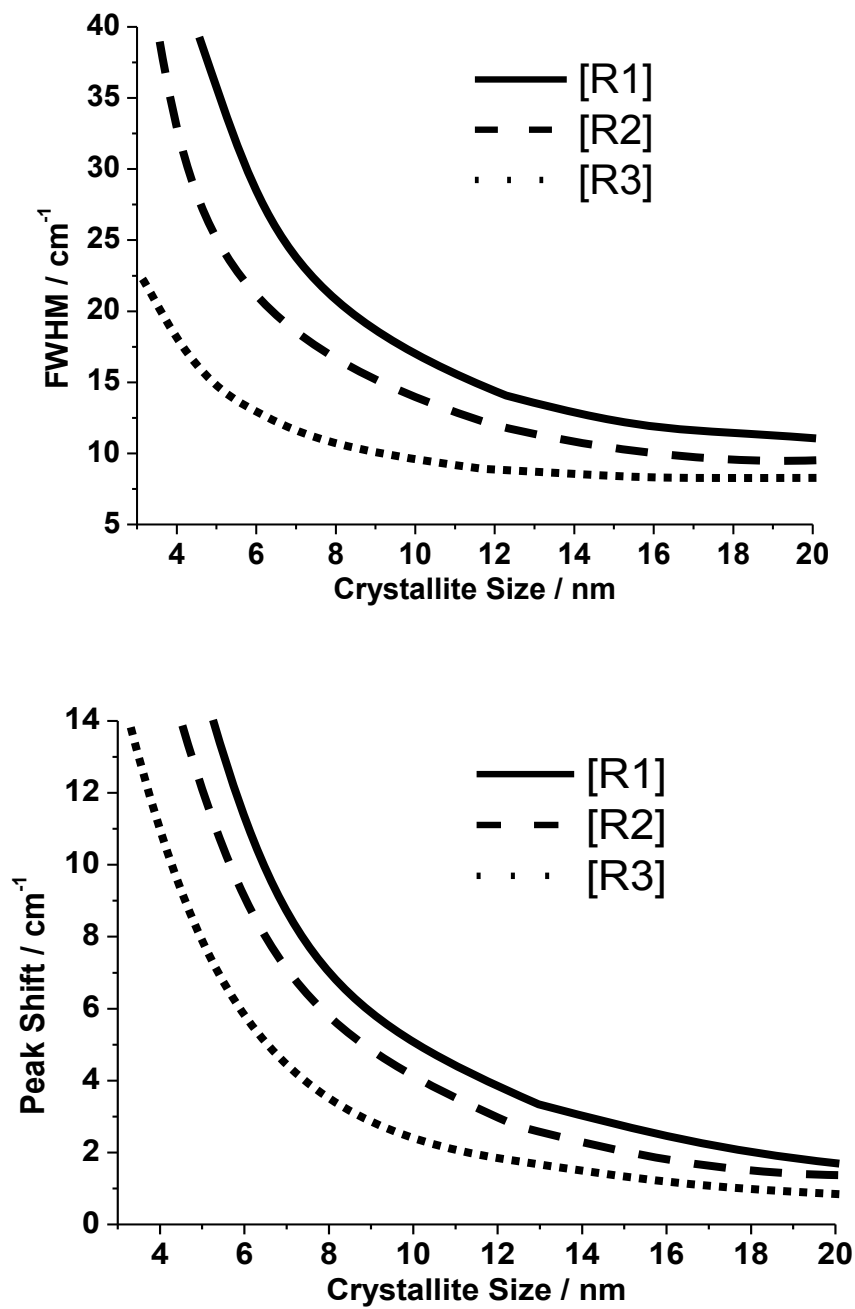

**Fig. S7** Various calculated models predicting the relationship of FWHM and peak shift of main Eg Raman line as a function of TiO$_2$ feature size.

[R1]: M. Ivanda , S. Music , M. Gotic , A. Turkovic , A. M. Tonejc , O. Gamulin , J. Mol. Struct. 1999 , 480 , 641.

[R2]: S. Balaji ,Y. Djaoued , J. Robichaud , J. Raman Spectrosc. 2006 , 37 , 1416 .

[R3]: D. Bersani , P. P. Lottici , X-Z. Ding , Appl. Phys. Lett. 1998 , 72 , 73.

**Rietveld analysis** of XRD spectra was carried out using the RIETAN-FP program [1] and Toroya's split pseudo-Voigt profile function for the calculations of structural parameters and integrated intensity. The diffraction patterns of the reference and H-implanted $TiO_2$ nanotubes were fitted by using a $TiO_2$ anatase model.

**$TiO_2$**

Structural model: Anatase $TiO_2$

Space group: $I\,4_1/a\,m\,d$ (VOL. A, 141)

$R_{WP}$ = 7.623%, $R_e$ = 4.164%

$a$ = 3.7877(2) Å, $c$ = 9.5090 (4) Å, $V$ = 136.42 (1) Å$^3$

| Atom | $x$ | $y$ | $z$ | $B$ |
|---|---|---|---|---|
| Ti | 0 | 0 | 0 | 1.07(8) |
| O | 0 | 0 | 0.2085(3) | 0.5(1) |

**H:$TiO_2$**

Structural model: Anatase $TiO_2$

Space group: $I\,4_1/a\,m\,d$ (VOL. A, 141)

$R_{WP}$ = 9.627%, $R_e$ = 6.057%

$a$ = 3.7904(6) Å, $c$ = 9.530(2) Å, $V$ = 136.93 (4) Å$^3$

| Atom | $x$ | $y$ | $z$ | $B$ |
|---|---|---|---|---|
| Ti | 0 | 0 | 0 | 1.2(3) |
| O | 0 | 0 | 0.204(2) | 0.8(5) |

Calculated intensity of reference and H-implanted nanotubes

| Index | $2\theta$ / ° | $d$ / Å | Integrated intensity Theoretical | Reference | H-implanted |
|---|---|---|---|---|---|
| 101 | 25.308 | 3.51629 | 100000 | 31223 | 27111 |
| 103 | 36.951 | 2.43073 | 6700 | 4294 | 1723 |
| 004 | 37.79 | 2.37865 | 20586 | 100000 | 7955 |
| 112 | 38.572 | 2.33222 | 7844 | 2466 | 2076 |
| 200 | 48.047 | 1.8921 | 29051 | 7032 | 6707 |
| 202 | 51.969 | 1.75815 | 0 | 0 | 0 |
| 105 | 53.885 | 1.70007 | 18812 | 21987 | 15665 |
| 211 | 55.072 | 1.66619 | 18637 | 4663 | 4124 |
| 213 | 62.117 | 1.49308 | 3353 | 924 | 621 |
| 204 | 62.692 | 1.48076 | 14761 | 6168 | 3432 |
| 116 | 68.756 | 1.3642 | 6674 | 5664 | 2370 |
| 220 | 70.303 | 1.33792 | 7278 | 1760 | 1507 |

The (001)-orientation degree was evaluated in terms of the Lotgering factor $F$(001), which is calculated from the following equation,[*2] $F = (P-P_0)/(1-P_0)$ where $P_0 = \Sigma I_0(hkl)/\Sigma I_0(HKL)$ and $P = \Sigma I(hkl)/\Sigma I(HKL)$. $I_0$ and $I$ are the integrated intensities of each of the diffraction peaks in X-ray diffraction patterns as presented in ICSD database and in experimental data, respectively. $F(001)$ values of the reference and H-implanted nanotubes were calculated to be 49.2% and 2.2%, respectively.

[*1] F. Izumi and K. Momma, *Solid State Phenom.*, 2007, **130**, 15.
[*2] F. K. Lotgering, *J. Inorg. Nucl. Chem.*, 1959, **9**, 113.